\documentclass[12pt,twocolumn,journal]{IEEEtran}

\usepackage{subfigure}
\usepackage{graphicx}
\usepackage{amsmath}
\usepackage{tabularx}
\usepackage{graphics}
\usepackage[dvips]{epsfig}

\newcommand{\hl}[1]{#1}
\begin{document}

\title{A Discussion on Context-awareness to Better Support the IoT Cloud/Edge Continuum}

\author{

   Daniel M. A. da Silva$^{1,3}$ and  Rute C. Sofia$^{2,3}$

\IEEEauthorblockA {$^{1}$ COPELABS - University Lusofona, Campo Grande 388, 1700-097 Lisboa, Portugal (E-mail: daniel.maniglia@hotmail.com)}

\IEEEauthorblockA{$^{2}$ Research Institute of the Free State Bavaria associated with the Technical University of Munich. Guerickerstr. 25, 80634 München (e-mail: sofia@fortiss.org)}

\IEEEauthorblockA{$^{3}$ University Lusofona, Campo Grande 376, 1700-097 Lisboa, Portugal}

}

\maketitle

\begin{abstract}
This paper debates on notions of context-awareness as a relevant asset of networking and computing architectures for an Internet of Things (IoT), in particular in regards to a smoother support of the the networking operation between Cloud and Edge. Specifically, the paper debates on notions of context-awareness and goes over different types of context-awareness indicators that are being applied to Edge selection algorithms, covering the approaches currently used, the role of the algorithms applied, their scope, and contemplated performance metrics. Lastly, the paper provides guidelines for future research in the context of Cloud-Edge and the application of context-awareness to assist in a higher degree of automation of the network and, as consequence, a better support of the Cloud to Edge continuum.
\end{abstract}

\section{Introduction}
\label{sec:introduction}
The daily routines of regular citizens integrate a wide variety of highly heterogeneous \textit{Internet of Things (IoT)} systems. Such systems integrate simple sensors and actuators, networking devices, and  more complex cyber-physical systems, such as smart sensors and mobile personal devices (e.g., smartphones) which further integrate a large number of sensing interfaces. For instance, in personal mobile devices, sensors such as accelerometer, GPS, microphone, or camera, bring in the possibility of exploiting new types of data coined as \textit{smart data} or \textit{small data}, derived from the track and trace process of different aspects of the routine of citizens, e.g., roaming habits; application usage; location preferences ~\cite{zheng2016monitoring,milovsevic2011applications,hansel2016wearable}.
While such sensorial capability is giving rise to new types of data and services, it  brings in additional computational and data exchange challenges. Firstly, the datasets are richer, even though data is fine-grained, and often polled more frequently, thus resulting in larger volumes of data to be analysed ~\cite{sun2016internet,kitchin2015small}.  
Secondly, the IoT communication architectural models that are being applied to support such data transmission cannot cope with the properties of such traffic (e.g., high volumes of small data packets). This is both due to the increasingly larger number of devices being interconnected to the Internet and to a higher heterogeneity of the hardware and software involved ~\cite{yaqoob2017internet}. \hl{Thirdly, the processing of the richer and more complex data sets require support from computationally heavy Artificial Intelligence (AI) engines supported by the Cloud.} While the Cloud helps in supporting the required data analytics complex computation, the more heterogeneous IoT scenarios available today are often not compatible with the delays derived from pushing all of the data processing and storage to the Cloud ~\cite{cozzolino2017enabling}. 

In the quest to assist smart data computation in IoT scenarios, related trends concern a decentralisation of Internet services and of networking functions across the so-called \textit{Cloud to Edge continuum (Cloud-Edge)}. The Cloud-Edge continuum refers to a set of operations that are required to fulfil, in an automated way, user and application requirements, taking into consideration networking features.  Today, the Cloud-Edge continuum relies already on context-awareness indicators, as shall be debated in section III and IV of the paper. However, this is limited, often tied to strict network guarantees, and such indicators are not sufficient, in our opinion, to sustain novel and more dynamic IoT environments, where the Edge is mobile, highly heterogeneous (e.g., an embedded device, a smart satellite).

Existing trends attempt to best serve mobility of devices and users; the need for data and user privacy; the larger volumes of sensitive data to be analysed, and the requirements to handle such data ~\cite{hosseinian2018emerging} ~\cite{sethi2017internet}. This is giving rise to alternative ways to provide data exchange in IoT environments, as occurs with the paradigm of Edge/Fog computing~\cite{bonomi2012fog}.
Usually, such paradigms take into consideration task, service and resource offloading, to assist in a better resource management. However, to support better dynamic environments, it is necessary to consider how to best adjust the computational needs to the respective context and hence, it is relevant to revisit notions of \textit{context-awareness}.

This is the motivation for this work. We believe that context-awareness can assist in a smoother transition of computational/storage/networking resources, from the Cloud to the Edges and vice-versa. To assist this debate, the paper contributions are three-fold: i) the paper provides a thorough review on work that focuses on context-awareness for IoT; ii) the paper contributes to the definition of context-awareness in IoT and debates on specific context-awareness indicators that can be considered to better support a smooth Cloud-Edge continuum; iii) the paper provides guidelines concerning the integration of context-awareness in Cloud-Edge IoT environments

The review provided in this paper has been based on an extensive review of papers concerning context-awareness for IoT environments. This review has comprised an analysis of papers from 2011 until 2020, based on the paper keywords "context", "context-awareness", "Edge computing", "behavior inference" and also focused on the area of "networking architectures", areas of interest of the authors. The selection took into consideration the following aspects: i)  the work has been described in peer-reviewed publications with a high Impact Factor; ii) most recent references have been preferred against older ones.

The paper is organised as follows. Section \ref{relatedwork} goes over related work, explaining the contributions of this paper towards related literature. Section \ref{iotbackground} describes background on IoT communication aspects, including notions for Edge/Fog computing. Section \ref{contextawareness} discusses the role of context-awareness for IoT and describes specific indicators that are being used to assist on a selection of the whereabouts to store and compute data. Section \ref{edgeselection} specifically focuses on the integration of context-awareness into IoT Fog/Edge architectures, detailing existing areas of interest.
Section \ref{conclusions} concludes the paper, discussing findings and providing a set of guidelines for future research.

\section{Related Work}

\label{relatedwork}
Several related work has focused on different categories of network communication challenges experienced in IoT scenarios. Specifically focusing on the domain of eHealth, Islam et al. describe on challenges existing in current IoT healthcare middleware~\cite{islam2015internet}.  Dimitrov et al. delve into issues concerning data mining, data storage, and data analysis~\cite{dimitrov2016medical} in IoT eHealth scenarios. Poon et al. focus on sensor communication~\cite{poon2015body}. Sensing and big data management have been debated by Yue Hong et al.~\cite{yuehong2016internet}, and the identification of key components of an end-to-end IoT has been discussed by Baker et al.~\cite{baker2017internet}. This line of work identifies and highlights challenges that IoT faces in Smart Health environments, including security, privacy, usability, energy awareness. This line of work is relevant to our work, given that eHealth scenarios experience specific challenges, in particular concerning data privacy and data sensitivity, challenges which can be lowered if the underlying networking architectures assist in handling data locally, within trusted environments.

In the context of IoT for Smart Cities environments, where smart applications are used to collect and to exchange different types of data, Sholl et al. propose  a  Smart City  architecture  that  harnesses  the  power  of semantic technologies to allow machines and people to understand the  relationships among data in a context-aware manner, and to extract knowledge~\cite{sholla2018semantic}. Choi et al. propose a software architecture to assist efficient middleware deployment in Smart Cities, by relying on semantic technologies~\cite{choi2018intelligent}.

Context-awareness is also highly relevant to data mining and classification as, for instance, debated in the context of vehicular networks by Ruta et al.~\cite{ruta2018knowledge}. 

Chen et al. surveyed Edge computing resource-efficient offloading mechanisms~\cite{chen2018thriftyedge}. Still in regards to Edge/Fog computation offloading, Wang et al.~\cite{wang2019context} collected and investigated key issues, methods related to the offloading problem in Cloud to Edge environments.

Another category of related work focuses on the understanding and definition of context and context-awareness, which are central points in this review paper. Some authors~\cite{brown1997context} define context in association with parameters such as location, neighbour identity, time-based indicators such as visit duration, environmental characteristics such as season, temperature. Ryan et al. define context as the user’s location, environment, identity, and the time~\cite{ryan1999enhanced}. Dey et al. states that context is the user’s emotional state, focus of attention, location and orientation, date and time, objects and people in the user’s environment~\cite{dey1998context}. Schilit et al. argue that the only important aspects of context are user location, the user’s neighbours, and resources near the user~\cite{schilit1994context}. They define context to be subject to the constantly changing execution environment and the environment is thus three-fold: computing environment, user environment and physical environment. Sofia et al.~\cite{sofia2017umobile} define context indicators based on the network layers, derived from roaming patterns of users.

Our work differs from the described related work in that it debates on research that applied context-awareness to assist in automating the IoT Cloud-Edge operation, surveying the use of context data to improve network performance in Edge/Fog Computing for environments exhibiting variability, such as occurs today in IoT environments that involve Thing-to-Thing and People-to-Thing interactions. 

\section{IoT Communication Background} 
\label{iotbackground}
IoT environments can be broadly grouped into two categories, related with the specific requirements and expected benefits: \textit{Consumer IoT (CIoT)} and \textit{Industrial IoT (IIoT)}~\cite{bandyopadhyay2011internet}~\cite{sofiar2019}. 
Both IoT categories rely on computational architectures that integrate four main functional blocks: data capture; data storage; data analysis; data exchange.
However, the requirements on these different environments introduce different challenges.

IIoT ~\cite{sisinni2018industrial} ~\cite{geng2017industrial} focuses on how smart machines, networked sensors, people, and data analytics can improve aspects such as productivity, service efficiency. IIoT is applied to different vertical markets, e.g., Industrial Automation, Smart Cities, Smart Factory, Logistics. Moreover, specific IIoT markets include also Smart Health, Smart Energy, or People-at-Work markets.

IIoT is expected to support both \textit{Machine to Machine (M2M)} and \textit{People-to-Machine} interaction, either for application monitoring, control, for instance, or as part of a self-organised system, with a distributed control which does not necessarily require human intervention. IIoT often implies higher data rates and larger data volumes. Moreover, applications are often mission and/or safety critical requiring strict and bounded guarantees, such as low delay, low jitter, or zero packet congestion.

CIoT concerns the use of IoT in aspects related to the daily living of people and aims at increasing usefulness of technology in such context. It involves scenarios focused on the interconnection of consumer and devices, as well as of anything involving the users' environments such as homes, offices, and cities~\cite{williams2017identifying}. Vertical markets of CIoT comprise, for instance, Smart Cities, Connected Mobility, Smart Health.

\textit{Personal IoT (PIoT)} is a sub-category of CIoT focused on the application of smart systems based on personal devices, as well as based on sets of sensors and actuators applied to improve quality of living. The most popular form of PIoT concerns fitness solutions aiming to bring awareness and to improve physical health of users~\cite{islam2015internet,yeole2016use}. Currently, these systems are more commonly used in the context of \textit{Ambient Assisted Living (AAL)}. AAL encompasses technical systems to support people with special needs in their daily routine, e.g.,  elderly ~\cite{mainetti2016iot}, temporarily disabled people,  or anyone that needs supportive monitoring. ~\cite{hassanalieragh2015health} ~\cite{ray2014home}. 

\subsection{Supporting Asynchronous and Many-to-many Communication}
\label{asynccommunication}
 From a protocol perspective, the interconnection of IoT Things and applications, be it directly to a controller or to the Cloud-Edge, has been traditionally deployed by having sensors harvesting information and sending such information to a specific device/system, for instance, an IoT gateway, an IoT broker. Hence, initially the point-to-point communication model provided by TCP/IP was enough to support the requirements of IoT data exchange. 
 
 With the increase of IoT devices, as well as with the new software-based and open-source approaches being explored, IoT services are becoming more complex, thus introducing additional requirements. Firstly, several, if not most of the devices in IoT scenarios are mobile. Secondly, the integration of the different hardware and software solutions that compose IoT environments is often provided by third-parties. Thirdly, IoT scenarios often accommodate hundreds or thousands of devices, often communicating across large distances.

To cope with these changes, data exchange in IoT needs to be supported by mechanisms capable of accommodating aspects such as mobility, security, large distances, intermittent connectivity. For this, it is necessary to support two main communication requirements: \textit{asynchronous communication support}, and
\textit{
many-to-many service distribution support}.

Internet communication protocols are therefore evolving, in the context of IoT, to support the 2 main mentioned requirements. For instance, the procotocols that support IoT data exchange (IP-based messaging protocols) usually rely on a \textit{broker-based} publish/subscribe communication model~\cite{sofiar2019}. The broker is a mediating functional entity that handles data being exchanged between producers and consumers in an asynchronous way. First, consumers subscribe specific data interests. Then they get the matching information provided by producers~\cite{naik2017choice}.
 Broker models create an abstraction layer as well, and can protect the identity of producers and subscribers. Nevertheless, they are still focused on reaching hosts (machines), and not really focused on the content.

The most recent evolution of  publish/subscriber models is embodied in the \textit{Information-centric Networking}  publish-subscriber paradigm ~\cite{zhang2014named}. \textit{Information-centric Network (ICN)} is a networking architectural paradigm that is focused on data reachability, instead of host reachability. In the context of IoT, ICN models seem to be promising as the network semantics that ICN automatically supports aspects such as consumer mobility~~\cite{fi11050111}, security, as well as address abstraction by design. There are today several ICN architectural proposals such as the \textit{Data-Oriented Network Architecture (DONA)}~\cite{koponen2007data}; the \textit{Network of information (NETINF)}~\cite{ahlgren2010second}; the \textit{ Content-Centric Networking (CCN)}~\cite{jacobson2009networking}; the \textit{Named Data Network (NDN)}~\cite{zhang2014named} .
Out of these, the networking architecture most suitable for IoT is the NDN architecture~\cite{meddeb2017information}.

The NDN architecture defines a simple and robust data-centric, pull-based and receiver-driven communication model based on the exchange of two packets types, \textit{Interest} and \textit{Data} packets. Interest packets are sent by consumers willing to express interest on specific content and contain hierarchical, global content names ~\cite{amadeo2014named}. Data packets are sent by producers upon the reception of Interest packets, and carry chunks of signed data.

\subsection{The Role of Edge/Fog Computing}
\label{edgefog}
Fog Computing~\cite{bonomi2012fog}, also known as Edge computing~\cite{shi2016edge}, extends the Cloud Computing paradigm to the "Edges" of the network, bringing in new opportunities to explore applications and services. By assisting the placement of storage and data processing closer to the data sources, Edge Computing brings in benefits in terms of latency and energy consumption~\cite{gedeon2019fog, sarddar2018refinement, iorga2018fog,choudhari2018prioritized,maiti2018qos}, for instance. 

In this paper Fog and Edge are used indistinctly, as we consider the most recent evolution of Edge, where the Edge is elastic in terms of whereabouts or even system composition (e.g., an Edge can be a smart sensor, a satellite, a smartphone, or an eNodeB)~\cite{yousefpour2019all}. However, other views provide a stricter perspective on Edge computing, derived from a telecommunications perspective. This is the case, for instance, of the Mobile Edge Computing (MEC) architecture, where the Edge is still within the control of the operator and consists of a specific computational unit, working in isolation or being complementar to the Cloud. While in Fog computing, the notion of Edge is more elastic, covering, for instance, field-level and end-user devices (e.g., smartphones, smart sensors)~\cite{chiang2017clarifying}.  

For IoT, and due to aspects such as security (e.g., the need to have in-plant security and resilient communication in IIoT scenarios), large distances, as well as large sets of frequent data lead to an insufficiency of the Cloud to satisfy the \textit{Quality of Service (QoS)} requirements (e.g., low latency) of different IoT applications.  Fog computing aims to overcome some limitations of Cloud-centric IoT-models by taking advantage of Edge network
resources~\cite{aazam2018deploying}. 

\begin{figure}[h]
    \centering
    \includegraphics[width=0.4\textwidth]{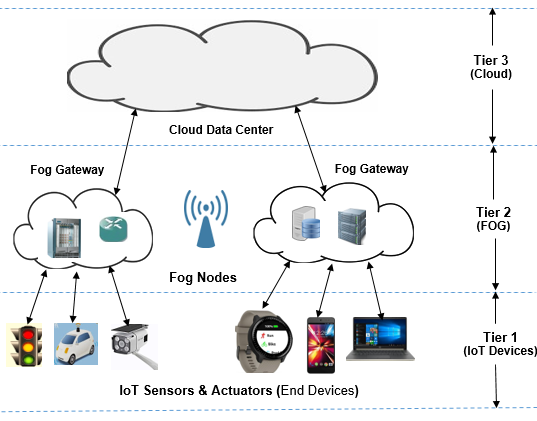}
    \caption{Fog/Cloud Computing Architecture.}
    \label{ARQUITETURA}
\end{figure}

Fog/Edge network architectures integrate mechanisms to better distribute data computation and data storage across a specific infrastructure.  Figure~\ref{ARQUITETURA} illustrates such a networking architecture, where Layers represent Tier levels. 

\textbf{Tier 1} integrates IoT \textit{field-level} devices, such as sensors and actuators. These are data sources,  devices that capture and distribute data to other Tier devices, same Tier, or next Tier level. \textbf{Tier 2} (FOG) integrates IoT devices coined as \textit{Fog nodes}  ~\cite{tordera2016fog}. IoT hubs and gateways that gather data and process information fall into this category. The Tier 2 level includes also devices such as routers and \textit{Access Points (AP)}. Fog nodes are arranged in a hierarchical way and communication is only possible between a parent-child pair in the hierarchy. Given that these devices are in the edges of the network, often located in Customer Premises, Fog nodes often have limited resources. \textbf{Tier 3} (CLOUD) devices often have a significantly higher amount of resources. These are, for instance, virtual machines in data centers.

\section{Context-awareness in IoT}
\label{contextawareness}
Context-aware computing has been used over the last decade in desktop  applications,  Web  applications, mobile computing, and  pervasive/ubiquitous computing. Context-aware computing is a computing paradigm in which applications can discover and take advantage of context information such as user location, time of day, neighbouring users and devices, user activity~\cite{musumba2013context}.  
Context is~\textit{"any information that can be used to characterise the  situation  of  an  entity.  An  entity  can be  a  person,  place,  or object that is considered relevant to the interaction between a user  and  an  application"}~~\cite{abowd1999towards}. 

Hence, there is a significant difference between context information and raw data sent by IoT devices. Raw data concerns unprocessed  data that is directly retrieved from data sources. Context information is generated by processing raw sensor data. Such data is validated, checked for consistency, and often annotated with meta-data~\cite{sanchez2006generic}. For  instance,  GPS sensor  readings can  be  considered  as  raw  sensor data. Once it represents a geographical location, it becomes context.

IoT environments comprise a large number of devices and large volumes of data to be transmitted and processed. Understanding how to use and how to process that data to generate relevant knowledge is therefore dependent on the type of context of services, users, as well as networking architectures. Hence, context-awareness plays a critical role in assisting decisions in terms of what data needs to be processed, where that data should be processed, and when. 

In regards to IoT environments, context-awareness is being applied to improve different computational aspects, as summarised in Table \ref{TableI}. The table categorizes related work first by area of application, explaining the purpose (column 2), and where the related work applies such improvements (column 3). The context-awareness indicators used are presented in column 4, while the applicability domain (vertical market) is provided in column 5. The related literature is placed in column 6.

\begin{table*}[]
\center
\caption{Context-awareness Application in IoT Computational and Networking Architectures.}
\label{TableI}
\begin{tabular}{|l|l|l|l|l|l|}
\hline
\multicolumn{1}{|c|}{ \textbf{Area}} & \multicolumn{1}{c|}{\textbf{Purpose}} & \multicolumn{1}{c|}{\textbf{Where}} & \multicolumn{1}{c|}{\textbf{Indicators}} & \multicolumn{1}{c|}{\begin{tabular}[c]{@{}l@{}}\textbf{Domain}\end{tabular}} & \multicolumn{1}{c|}{\textbf{References}} \\ \hline
\begin{tabular}[c]{@{}l@{}}Authentication\\ and Control\end{tabular} & \begin{tabular}[c]{@{}l@{}}To facilitate secure \\ authentication\\ and control of  IoT  devices  in\\ untrusted environments\end{tabular} & \begin{tabular}[c]{@{}l@{}}End-user, Edge, \\ Cloud\end{tabular} & \begin{tabular}[c]{@{}l@{}}Physical context\\ (light, temperature, noise);\\ computing context\\ (app usage, touch\\ patterns), \\ User context  (roaming\\ patterns, neighbor  cluster,\\ location, etc.)\end{tabular} & \begin{tabular}[c]{@{}l@{}}PIoT, SmartHome,\\ SmartHealth, \\ Automation\end{tabular} & ~\cite{kaur2018authentication,reddy2019context,wang2019context,loske2019context,kausar2019mutually,anton2019putting,trnka2018using,smith2018context,habib2015context,doukas2012bringing,zhang2016sharing} \\ \hline 
\begin{tabular}[c]{@{}l@{}}Resource\\ management\\ and orchestration\end{tabular} & \begin{tabular}[c]{@{}l@{}}To improve the overall IoT \\ computational platform and \\ services providing control \\ based on context\end{tabular} & Edge, Cloud & \begin{tabular}[c]{@{}l@{}}Device usage and\\ resources, time, location,\\ user behavior\end{tabular} & \begin{tabular}[c]{@{}l@{}} SmartHome, \\ SmartCities \end{tabular}& ~\cite{garcia2018air4people, dobrescu2019context, bacstuug2016proactive,zeydan2016big,zhao2018collaborative,lee2018hierarchical,afzal2017energy,flores2018evidence,tang2018multi} \\ \hline
Forwarding/routing & \begin{tabular}[c]{@{}l@{}}To introduce context-awareness \\to the network \\ layers in order to provide better \\ chances to forward, in particular \\ in highly heterogeneous scenarios\end{tabular} & \begin{tabular}[c]{@{}l@{}}End-user, Edge, \\ Cloud\end{tabular} & \begin{tabular}[c]{@{}l@{}}Location, roaming data, \\ accelerometer, speed, \\ device usage (e.g., battery)\end{tabular} & \begin{tabular}[c]{@{}l@{}}Opportunistic IoT \\ environments, \\ SmartCities, PIoT\end{tabular} & ~\cite{sofia2017umobile,dhumane2018context,guo2013opportunistic,musolesi2009car,dhumane2016routing} \\ \hline
Offloading & \begin{tabular}[c]{@{}l@{}}To assist in deciding where \\ to store data and to compute \end{tabular} & Edge, Cloud & \begin{tabular}[c]{@{}l@{}}Location, battery, latency \\ and other network \\ measurement indicators\end{tabular} & \begin{tabular}[c]{@{}l@{}}Smart Logistics, \\ SmartCities, \\ SmartMobility\end{tabular} & ~\cite{tran2016collaborative,lee2018hierarchical,pham2019joint,tripathi2017adaptive,zhang2019deep} \\ \hline
\begin{tabular}[c]{@{}l@{}}Semantic \\ interoperability \end{tabular} & \begin{tabular}[c]{@{}l@{}}To assist in a \\ smoother operation in large-scale \\ heterogeneous environments\end{tabular} & Edge & \begin{tabular}[c]{@{}l@{}}Application delay \\ requirements, request \\ history, service/user \\ similarity\end{tabular} & \begin{tabular}[c]{@{}l@{}}Smart Logistics, \\ SmartCities, \\ SmartMobility\end{tabular} & ~\cite{tu2017cane,hu2014multidimensional,alirezaie2017ontology,mingozzi2016semantic,fortino2017modeling, casadei2019development} 
\\  \hline
\begin{tabular}[c]{@{}l@{}} Multi-layer \\ interoperability \end{tabular}& \begin{tabular}[c]{@{}l@{}}To provide IoT interoperable \\ ecosystems by using \\ context-awareness \\ in a bottom-up approach\end{tabular} & Edge, Cloud & \begin{tabular}[c]{@{}l@{}}  Users, \\device usage, \\environmental\\ indicators \end{tabular} & \begin{tabular}[c]{@{}l@{}}Data discovery, \\ management   and \\ communication\end{tabular} & ~\cite{fortino2018towards,aloi2017enabling,fortino2017modeling,casadei2019modelling,casadei2019development}

\\ \hline
\end{tabular}\end{table*}

A first area of related work \hl{(row 1)} applies context-awareness to authentication and control in untrusted environments. Context-aware access control mechanisms are being used to provide system access, using the user personal data context and not personal data.   

A second area of related work \hl{(row 2)} concerns the application of context-awareness for resource management and orchestration. Such line of work focuses on improving the overall computational and networking performance by exploring context-awareness to reduce energy consumption; reduce overall latency; message overload. Context-awareness is relevant to assist in deciding when and where to process data, thus contributing to latency reduction, for instance~\cite{tran2016collaborative,zeydan2016big}.

In regards to forwarding/routing applications \hl{(row 3)}, one example of the work being pursued is to take into consideration, at a network level, the context that surrounds users and that can assist in better defining opportunities for data transmission over time, and space, i.e., context-awareness at the network layers~\cite{sofia2017umobile}. Context-awareness can also assist in a better distribution of in-network caching; more efficient naming aggregation, as well as in a more efficient data transmission in the context of large-scale scenarios~\cite{bacstuug2016proactive,chang2018learn}. 

\hl{Another category of work focuses on applying context-awareness to offloading}\hl{(row 4)}, \hl{i.e., to decide where to store data, and also where to compute such data. For this purpose, parameters such as location, residual energy of the device are being applied.}

\hl{A fifth category of related work focuses on semantic interoperability aspects, including related work that has been delving on improving data sharing on upper layers via semantic modelling. Once the information can be collected from a range of sources and some information must be explicitly supplied by users, context-awareness can be applied to identify  the relationships level between people, the ownership of devices and communication channels providing a seamless approach to the interconnection of devices and their data exchange, by providing automated support to the interconnection of, for instance, different data models derived from different applicability domains}\hl{(row 5). In this context, indicators derived from the application layer (such as delay requirements), or even similarity between used services is being applied to assist in an automated interconnection}.

\hl{The last row (row 6) covers work related with multi-layer interoperability. This work focuses on discovery, management and high-level communication of IoT devices in heterogeneous IoT platforms, defining, for instance, component-based methods for middleware interoperability.} 

\section{Context-awareness and Selection Algorithms}
\label{edgeselection}
Edge selection algorithms provide a smoother operation in Cloud-Edge environments, in particular when considering services and applications that might require very short response times, or applications that might produce a large quantity of data to be processed. Sending such data to the Cloud may result in large delays, or excessive energy consumption by the network devices, for instance. An example of technological solutions that require adaptation on the go are \textit{Mobile Pervasive Augmented Reality (MPARS)}~\cite{pascoal2020mobile}. As stated by Pascoal et al., context-awareness derived from the surrounding environment, as well as from the user's habits, and computational preferences can assist a better aggregation and placement of data. This also assists in extending the reach of computational and networking architectures, considering Edges that are mobile and resource constrained.

Current Edge placement algorithms are often focused on aspects such as latency and energy improvement, as summarised in Table~\ref{algorithms}, \hl{which summarises Edge selection algorithms, categorizing them by context information considered (column 2), scope (column 3), as well as performance metrics relied by the algorithm (column 4).}

\begin{table*}[]
\center
\caption{Edge Selection Algorithms.}
\label{algorithms}
\begin{tabular}{|p{1cm}|p{3cm}|p{4cm}|p{4cm}|p{3cm}|p{3cm}|}
\hline
\multicolumn{1}{|c|}{\textbf{Nr.}}  &\multicolumn{1}{|c|}{\textbf{Algorithm}}  & \multicolumn{1}{c|}{\textbf{Context}} & \multicolumn{1}{c|}{\textbf{Scope}} & \multicolumn{1}{c|}{\textbf{Metrics}} \\ \hline
1 &\begin{tabular}[c]{@{}l@{}} XTC~\cite{wattenhofer2004xtc}\end{tabular} & Neighbor ordering, neighbor order exchange, Edge selection (no need for node location or global topology knowledge) & \begin{tabular}[c]{@{}l@{}}Network control \\ algorithm for ad-hoc environments, \\ node selection\end{tabular} & \begin{tabular}[c]{@{}l@{}}Latency\end{tabular} \\ \hline
2 & AR Edge Selection ~\cite{maheshwari2018scalability} & \begin{tabular}[c]{@{}l@{}}Application requirements \\ and traffic load\end{tabular} & Edge Selection & \begin{tabular}[c]{@{}l@{}}Latency threshold\end{tabular} \\ \hline
3 & \begin{tabular}[c]{@{}l@{}}Opportunistic \\ Routing (ENS\_OR)~\cite{luo2015opportunistic}\end{tabular}  & \begin{tabular}[c]{@{}l@{}}Distance to sink \\ and residual energy \\ on both neighboring nodes\end{tabular} & Opportunistic relaying algorithm, node selection & \begin{tabular}[c]{@{}l@{}}Distance to sink \\ and residual energy \end{tabular} \\ \hline
4 & RNST ~\cite{han2009reference} & Node location via trilateration  & \begin{tabular}[c]{@{}l@{}}Wireless indoor location, \\ node selection\end{tabular} & \begin{tabular}[c]{@{}l@{}}relative direct distance\\ radio range\end{tabular} \\  \hline
5 & Latency Bounded Minimum Influential Node Selection~\cite{zou2009latency}  & \begin{tabular}[c]{@{}l@{}}Diffusion, influential nodes impact \\ on the speed of diffusion\end{tabular} & \begin{tabular}[c]{@{}l@{}}Node Selection \end{tabular} & \begin{tabular}[c]{@{}l@{}}Propagation speed \\ of influential nodes\end{tabular} \\ \hline
6 & \begin{tabular}[c]{@{}l@{}} Computation \\ and networking load \\ node selection ~\cite{subhlok1999automatic} \end{tabular}  & \begin{tabular}[c]{@{}l@{}}Application requirements \\ node resources \\ link quality resources\end{tabular} & \begin{tabular}[c]{@{}l@{}}Node Selection\end{tabular} & \begin{tabular}[c]{@{}l@{}} Node and link availability\end{tabular} \\ \hline
7 & Branch-and-bound algorithm ~\cite{pham2019joint}  & \begin{tabular}[c]{@{}l@{}}Computational overhead\end{tabular} & \begin{tabular}[c]{@{}l@{}}Joint node \\ selection and \\ resource allocation\end{tabular} & \begin{tabular}[c]{@{}l@{}}Computation \\ overhead threshold\end{tabular} \\ \hline
8 & Threshold based policy~\cite{zhao2015cooperative}  &   \begin{tabular}[c]{@{}l@{}} Network policies \end{tabular} & Edge/Cloud selection & \begin{tabular}[c]{@{}l@{}}Delay threshold\end{tabular} \\ \hline
9 & ThriftyEdge~\cite{chen2018thriftyedge} & Task resource usage &   \begin{tabular}[c]{@{}l@{}}  Resource-efficient computation \\ Node selection \end{tabular}  & \begin{tabular}[c]{@{}l@{}} Latency, \\  minimum  \\ node usage  \\ (e.g., CPU, memory) \end{tabular} \\ \hline

10 & Edge Selected MPA ~\cite{wang2016edge} & Topology, neighborhood status  &   \begin{tabular}[c]{@{}l@{}}  Edge Selection \end{tabular}  & \begin{tabular}[c]{@{}l@{}} Channel quality \\ and proximity \end{tabular} \\ \hline

\end{tabular}
\end{table*}

Wattenhofer and Zollinger~\cite{wattenhofer2004xtc} propose XTC (1), a topology control algorithm to select the nearest Edge in ad-hoc wireless networks. The algorithm has three steps: 1- Neighbour ordering; 2- neighbour order exchange, and 3- Edge selection. It also has the advantage of not requiring full knowledge of the topology, or prior status on the node whereabouts. It therefore applies heuristics that take into consideration the direct neighborhood of the node, at different instant in times.

Sumit et al.,  propose an \textit{Edge selection algorithm for AR} applications (2)~\cite{maheshwari2018scalability}. Their algorithm takes into consideration both application requirements and traffic load. The algorithm scans the state of neighboring edges to find a "best" Edge which can serve the user within a specified latency threshold. 

An \textit{Energy Saving via Opportunistic Routing algorithm} (3) is proposed by Luo et al.~\cite{luo2015opportunistic}. This algorithm is applied in wireless networks and has two steps to select nodes: 1- selects a set of nodes with higher centrality; and 2- considers the status provided by other nodes it encounters. In terms of indicators, it considers the node's distance to the data sink, and the residual energy on both the parent and successor nodes.

The \textit{RNST} algorithm (4)~\cite{han2009reference} was developed to support mobile nodes for indoor wireless networks. It provides node location via trilateration, considering four steps: 1- A mobile node broadcasts a location message to its neighboring reference nodes, then the reference nodes return a confirmed location message; 2- The mobile node calculates the distances between each pair of nodes and judges if any of the three reference nodes can form almost equilateral triangle; 3- Compute the estimated locations of the mobile node using each of the possible equilateral triangles; 4- The mobile node calculates the average location value.

A \textit{Latency-bounded Minimum Influential Node Selection Algorithm} (5), proposed by Zou et al.~\cite{zou2009latency}, provides a selection of the most influential nodes on a (social) network, where most influential relates with the speed of diffusion, and not with connectivity. The algorithm steps are: 1- find a 1-hop dominating set for the rest of the nodes that are INACTIVE 2- the vertices that could be influenced by the 1-hop Latency-Bounded Minimum Influential Node Selection in Social Networks. 

A computation and networking load node selection algorithm (6)~\cite{subhlok1999automatic} is one of the first works, to our knowledge, that realises the need to meet, in an integrated way, both application and networking requirements. It relies on node resources such as CPU, and link resources, and considers as selection metrics node availability derived from a node and link QoS perspective.

The \textit{branch-and-bound algorithm} (7) ~\cite{pham2019joint} proposed by Pham et al. addresses both node selection and resource allocation. It is a Divide and conquer algorithm that uses computation overhead to select a node. 

Zhao et al. provide a threshold-based policy mechanism (8)~\cite{zhao2015cooperative} which finds an optimal local node to run delay-tolerant applications in mobile Cloud computing, designing a scheduling scheme to realize the cooperation between the local Cloud and the Internet Cloud. 

Xu Chen et al. introduce \textit{ThriftyEdge} (9)~\cite{chen2018thriftyedge}, a resource-efficient IoT task offloading algorithm. The authors rely on a hybrid approach to exploit the hierarchical resources across local nodes, nearby helper nodes, and the Edge-Cloud in proximity. They propose a topology-sorting-based task graph partition algorithm in order to reduce the Edge resource occupancy (usage). 

Yudan Wang and Ling Qiu propose \textit{MPA (ES-MPA)} (10)~\cite{wang2016edge}, a low complexity Edge discovery and selection approach to better support the massive connectivity of cellular IoT. 

Summarising, most of the existing algorithms that provide support for node selection usually consider a minor set of network or node requirements, e.g., latency, residual energy. Less common is the attempt to combine application/task and network requirements. Moreover, out of the analysis performed, we did not find algorithms that took into consideration behavior inference (node, link, service, and user), for instance.


\section{Conclusions and Guidelines for Future Research} \label{conclusions}
This paper reviews work concerning the relevancy of integrating context-awareness to improve the IoT data exchange across Edge and Cloud, in particular regarding the needs of IoT services and applications. The paper provides an overview on the needs of different IoT environments and revises proposals which consider context-awareness indicators to provide operational improvements, e.g., latency reduction, lower energy consumption. 
The review shows that the role of context-awareness in IoT environments is acknowledged, but that its integration to support more dynamic IoT environments is still limited, often being defined simply as location to assist traffic locality, or node resources, as described in section \ref{contextawareness}. As also debated in section \ref{contextawareness}, there are several opportunities to improve the Edge-Cloud continuum, by considering different levels of context-awareness indicators, derived from application requirements and from networking requirements, and also derived from the behaviour learning of inference of user activities and habits (e.g., roaming patterns; preferred network locations).
It is therefore relevant to consider some of the findings, to derive guidelines for future research. A summary of such guidelines is:

\begin{itemize}
      
    \item IoT applications are becoming more and more distributed across the Cloud and Edge, as addressed in section \ref{edgefog}. Edge selection mechanisms (cf. section \ref{edgeselection}) consider a limited integration of context-awareness. Other indicators which may better support more dynamic environments (e.g., indicators derived from mobility patterns) can be considered, thus being a relevant area of future work.
    \item The support of many-to-many asynchronous communication is today based on publish/subscribe models, as described in section \ref{asynccommunication}, is relevant to better support the needs of IoT data exchange. It provides the opportunity to scale better in comparison to the traditional client/server communication models, through parallel operation, message caching, tree-based or network-based routing. In addition to the IP-based messaging protocols commonly used in IoT environments, it is relevant to further delve on the relevancy of paradigms such as ICN, and focus on the integration of ICN architectures, such as NDN, into IoT. A relevant research area, which has been initiated but still requires much more exploration be it in terms of performance measurement or in terms of network architectures evolution is the applicability of ICN paradigms into IoT environments. 
    \item Variable and heterogeneous IoT scenarios, such as the ones embodied in PIoT, will benefit from bringing data processing closer to the end-user, as discussed in section \ref{edgeselection}. Context-awareness therefore plays a relevant role, be it in terms of better defining traffic and computational locality, or to assist in a more automated behavior of IoT networking architectures, end-to-end.
    \item To promote feedback in close-to-realtime, context-awareness can assist the network in making decisions that improve the network operation and, as consequence, can also improve data processing. 
\end{itemize}




\bibliographystyle{IEEEtran}

\bibliography{bibtex}

\begin{thebibliography}{100}
\providecommand{\url}[1]{#1}
\csname url@samestyle\endcsname
\providecommand{\newblock}{\relax}
\providecommand{\bibinfo}[2]{#2}
\providecommand{\BIBentrySTDinterwordspacing}{\spaceskip=0pt\relax}
\providecommand{\BIBentryALTinterwordstretchfactor}{4}
\providecommand{\BIBentryALTinterwordspacing}{\spaceskip=\fontdimen2\font plus
\BIBentryALTinterwordstretchfactor\fontdimen3\font minus
  \fontdimen4\font\relax}
\providecommand{\BIBforeignlanguage}[2]{{%
\expandafter\ifx\csname l@#1\endcsname\relax
\typeout{** WARNING: IEEEtran.bst: No hyphenation pattern has been}%
\typeout{** loaded for the language `#1'. Using the pattern for}%
\typeout{** the default language instead.}%
\else
\language=\csname l@#1\endcsname
\fi
#2}}
\providecommand{\BIBdecl}{\relax}
\BIBdecl

\bibitem{zheng2016monitoring}
H.~Zheng and J.~Jumadinova, ``Monitoring the well-being of a person using a
  robotic-sensor framework,'' in \emph{2016 AAAI Spring Symposium
  Series}.\hskip 1em plus 0.5em minus 0.4em\relax Association for the
  Advancement of Artificial Intelligence, 2016.

\bibitem{milovsevic2011applications}
M.~Milo{\v{s}}evi{\'c}, M.~T. Shrove, and E.~Jovanov, ``Applications of
  smartphones for ubiquitous health monitoring and wellbeing management,''
  \emph{JITA-JOURNAL OF INFORMATION TECHNOLOGY AND APPLICATIONS}, vol.~1,
  no.~1, 2011.

\bibitem{hansel2016wearable}
K.~H{\"a}nsel, ``Wearable and ambient sensing for well-being and emotional
  awareness in the smart workplace,'' in \emph{Proceedings of the 2016 ACM
  International Joint Conference on Pervasive and Ubiquitous Computing:
  Adjunct}.\hskip 1em plus 0.5em minus 0.4em\relax ACM, 2016, pp. 411--416.

\bibitem{sun2016internet}
Y.~Sun, H.~Song, A.~J. Jara, and R.~Bie, ``Internet of things and big data
  analytics for smart and connected communities,'' \emph{IEEE access}, vol.~4,
  pp. 766--773, 2016.

\bibitem{kitchin2015small}
R.~Kitchin and T.~P. Lauriault, ``Small data in the era of big data,''
  \emph{GeoJournal}, vol.~80, no.~4, pp. 463--475, 2015.

\bibitem{yaqoob2017internet}
I.~Yaqoob, E.~Ahmed, I.~A.~T. Hashem, A.~I.~A. Ahmed, A.~Gani, M.~Imran, and
  M.~Guizani, ``Internet of things architecture: Recent advances, taxonomy,
  requirements, and open challenges,'' \emph{IEEE wireless communications},
  vol.~24, no.~3, pp. 10--16, 2017.

\bibitem{cozzolino2017enabling}
V.~Cozzolino, A.~Y. Ding, J.~Ott, and D.~Kutscher, ``Enabling fine-grained edge
  offloading for iot,'' in \emph{Proceedings of the SIGCOMM Posters and
  Demos}.\hskip 1em plus 0.5em minus 0.4em\relax ACM, 2017, pp. 124--126.

\bibitem{hosseinian2018emerging}
A.~Hosseinian-Far, M.~Ramachandran, and C.~L. Slack, ``Emerging trends in cloud
  computing, big data, fog computing, iot and smart living,'' in
  \emph{Technology for Smart Futures}.\hskip 1em plus 0.5em minus 0.4em\relax
  Springer, 2018, pp. 29--40.

\bibitem{sethi2017internet}
P.~Sethi and S.~R. Sarangi, ``Internet of things: architectures, protocols, and
  applications,'' \emph{Journal of Electrical and Computer Engineering}, vol.
  2017, 2017.

\bibitem{bonomi2012fog}
F.~Bonomi, R.~Milito, J.~Zhu, and S.~Addepalli, ``Fog computing and its role in
  the internet of things,'' in \emph{Proceedings of the first edition of the
  MCC workshop on Mobile cloud computing}.\hskip 1em plus 0.5em minus
  0.4em\relax ACM, 2012, pp. 13--16.

\bibitem{islam2015internet}
S.~R. Islam, D.~Kwak, M.~H. Kabir, M.~Hossain, and K.-S. Kwak, ``The internet
  of things for health care: a comprehensive survey,'' \emph{IEEE Access},
  vol.~3, pp. 678--708, 2015.

\bibitem{dimitrov2016medical}
D.~V. Dimitrov, ``Medical internet of things and big data in healthcare,''
  \emph{Healthcare informatics research}, vol.~22, no.~3, pp. 156--163, 2016.

\bibitem{poon2015body}
C.~C. Poon, B.~P. Lo, M.~R. Yuce, A.~Alomainy, and Y.~Hao, ``Body sensor
  networks: In the era of big data and beyond,'' \emph{IEEE reviews in
  biomedical engineering}, vol.~8, pp. 4--16, 2015.

\bibitem{yuehong2016internet}
Y.~Yuehong, Y.~Zeng, X.~Chen, and Y.~Fan, ``The internet of things in
  healthcare: An overview,'' \emph{Journal of Industrial Information
  Integration}, vol.~1, pp. 3--13, 2016.

\bibitem{baker2017internet}
S.~B. Baker, W.~Xiang, and I.~Atkinson, ``Internet of things for smart
  healthcare: Technologies, challenges, and opportunities,'' \emph{IEEE
  Access}, vol.~5, pp. 26\,521--26\,544, 2017.

\bibitem{sholla2018semantic}
S.~Sholla, R.~Naaz, and M.~A. Chishti, ``Semantic smart city: Context aware
  application architecture,'' in \emph{2018 Second International Conference on
  Electronics, Communication and Aerospace Technology (ICECA)}.\hskip 1em plus
  0.5em minus 0.4em\relax IEEE, 2018, pp. 721--724.

\bibitem{choi2018intelligent}
C.~Choi, C.~Esposito, H.~Wang, Z.~Liu, and J.~Choi, ``Intelligent power
  equipment management based on distributed context-aware inference in smart
  cities,'' \emph{IEEE Communications Magazine}, vol.~56, no.~7, pp. 212--217,
  2018.

\bibitem{ruta2018knowledge}
M.~Ruta, F.~Scioscia, F.~Gramegna, S.~Ieva, E.~Di~Sciascio, and R.~P. De~Vera,
  ``A knowledge fusion approach for context awareness in vehicular networks,''
  \emph{IEEE Internet of Things Journal}, vol.~5, no.~4, pp. 2407--2419, 2018.

\bibitem{chen2018thriftyedge}
X.~Chen, Q.~Shi, L.~Yang, and J.~Xu, ``Thriftyedge: Resource-efficient edge
  computing for intelligent iot applications,'' \emph{IEEE network}, vol.~32,
  no.~1, pp. 61--65, 2018.

\bibitem{wang2019context}
R.~Wang and D.~Tao, ``Context-aware implicit authentication of smartphone users
  based on multi-sensor behavior,'' \emph{IEEE Access}, vol.~7, pp.
  119\,654--119\,667, 2019.

\bibitem{brown1997context}
P.~J. Brown, J.~D. Bovey, and X.~Chen, ``Context-aware applications: from the
  laboratory to the marketplace,'' \emph{IEEE personal communications}, vol.~4,
  no.~5, pp. 58--64, 1997.

\bibitem{ryan1999enhanced}
N.~Ryan, J.~Pascoe, and D.~Morse, ``Enhanced reality fieldwork: the context
  aware archaeological assistant,'' \emph{Bar International Series}, vol. 750,
  pp. 269--274, 1999.

\bibitem{dey1998context}
A.~K. Dey, ``Context-aware computing: The cyberdesk project,'' in
  \emph{Proceedings of the AAAI 1998 Spring Symposium on Intelligent
  Environments}, 1998, pp. 51--54.

\bibitem{schilit1994context}
B.~Schilit, N.~Adams, and R.~Want, ``Context-aware computing applications,'' in
  \emph{Mobile Computing Systems and Applications, 1994. Proceedings., Workshop
  on}.\hskip 1em plus 0.5em minus 0.4em\relax IEEE, 1994, pp. 85--90.

\bibitem{sofia2017umobile}
R.~C. Sofia, I.~Santos, J.~Soares, S.~Diamantopoulos, C.-A. Sarros,
  D.~Vardalis, V.~Tsaoussidis, and A.~d'Angelo, ``Umobile d4. 5: Report on data
  collection and inference models,'' UMOBILE Consortium, Tech. Rep., 2017.

\bibitem{bandyopadhyay2011internet}
D.~Bandyopadhyay and J.~Sen, ``Internet of things: Applications and challenges
  in technology and standardization,'' \emph{Wireless Personal Communications},
  vol.~58, no.~1, pp. 49--69, 2011.

\bibitem{sofiar2019}
R.~C. Sofia, ``An overview on the evolution of iot communication approaches
  (pre-print),'' \emph{MDPI pre-print}, 2019.

\bibitem{sisinni2018industrial}
E.~Sisinni, A.~Saifullah, S.~Han, U.~Jennehag, and M.~Gidlund, ``Industrial
  internet of things: Challenges, opportunities, and directions,'' \emph{IEEE
  Transactions on Industrial Informatics}, vol.~14, no.~11, pp. 4724--4734,
  2018.

\bibitem{geng2017industrial}
H.~Geng, ``The industrial internet of things (iiot),'' 2017.

\bibitem{williams2017identifying}
R.~Williams, E.~McMahon, S.~Samtani, M.~Patton, and H.~Chen, ``Identifying
  vulnerabilities of consumer internet of things (iot) devices: A scalable
  approach,'' in \emph{2017 IEEE International Conference on Intelligence and
  Security Informatics (ISI)}.\hskip 1em plus 0.5em minus 0.4em\relax IEEE,
  2017, pp. 179--181.

\bibitem{yeole2016use}
A.~S. Yeole and D.~Kalbande, ``Use of internet of things (iot) in healthcare: A
  survey,'' in \emph{Proceedings of the ACM Symposium on Women in Research
  2016}.\hskip 1em plus 0.5em minus 0.4em\relax ACM, 2016, pp. 71--76.

\bibitem{mainetti2016iot}
L.~Mainetti, L.~Patrono, A.~Secco, and I.~Sergi, ``An iot-aware aal system for
  elderly people,'' in \emph{Computer and Energy Science (SpliTech),
  International Multidisciplinary Conference on}.\hskip 1em plus 0.5em minus
  0.4em\relax IEEE, 2016, pp. 1--6.

\bibitem{hassanalieragh2015health}
M.~Hassanalieragh, A.~Page, T.~Soyata, G.~Sharma, M.~Aktas, G.~Mateos,
  B.~Kantarci, and S.~Andreescu, ``Health monitoring and management using
  internet-of-things (iot) sensing with cloud-based processing: Opportunities
  and challenges,'' in \emph{2015 IEEE international conference on services
  computing (SCC)}.\hskip 1em plus 0.5em minus 0.4em\relax IEEE, 2015, pp.
  285--292.

\bibitem{ray2014home}
P.~P. Ray, ``Home health hub internet of things (h 3 iot): An architectural
  framework for monitoring health of elderly people,'' in \emph{Science
  Engineering and Management Research (ICSEMR), 2014 International Conference
  on}.\hskip 1em plus 0.5em minus 0.4em\relax IEEE, 2014, pp. 1--3.

\bibitem{naik2017choice}
N.~Naik, ``Choice of effective messaging protocols for iot systems: Mqtt, coap,
  amqp and http,'' in \emph{2017 IEEE international systems engineering
  symposium (ISSE)}.\hskip 1em plus 0.5em minus 0.4em\relax IEEE, 2017, pp.
  1--7.

\bibitem{zhang2014named}
L.~Zhang, A.~Afanasyev, J.~Burke, V.~Jacobson, P.~Crowley, C.~Papadopoulos,
  L.~Wang, B.~Zhang \emph{et~al.}, ``Named data networking,'' \emph{ACM SIGCOMM
  Computer Communication Review}, vol.~44, no.~3, pp. 66--73, 2014.

\bibitem{fi11050111}
\BIBentryALTinterwordspacing
R.~C.~Sofia, ``Guidelines towards information-driven mobility management,''
  \emph{Future Internet}, vol.~11, no.~5, 2019. [Online]. Available:
  \url{https://www.mdpi.com/1999-5903/11/5/111}
\BIBentrySTDinterwordspacing

\bibitem{koponen2007data}
T.~Koponen, M.~Chawla, B.-G. Chun, A.~Ermolinskiy, K.~H. Kim, S.~Shenker, and
  I.~Stoica, ``A data-oriented (and beyond) network architecture,'' in
  \emph{ACM SIGCOMM Computer Communication Review}, vol.~37, no.~4.\hskip 1em
  plus 0.5em minus 0.4em\relax ACM, 2007, pp. 181--192.

\bibitem{ahlgren2010second}
B.~Ahlgren, M.~D’ambrosio, C.~Dannewitz, A.~Eriksson, J.~Golic,
  B.~Gr{\"o}nvall, D.~Horne, A.~Lindgren, O.~M{\"a}mmel{\"a}, M.~Marchisio
  \emph{et~al.}, ``Second netinf architecture description,'' \emph{4WARD EU FP7
  Project, Deliverable D-6.2 v2. 0}, 2010.

\bibitem{jacobson2009networking}
V.~Jacobson, D.~K. Smetters, J.~D. Thornton, M.~F. Plass, N.~H. Briggs, and
  R.~L. Braynard, ``Networking named content,'' in \emph{Proceedings of the 5th
  international conference on Emerging networking experiments and
  technologies}.\hskip 1em plus 0.5em minus 0.4em\relax ACM, 2009, pp. 1--12.

\bibitem{meddeb2017information}
M.~Meddeb, ``Information-centric networking, a natural design for iot
  applications?'' Ph.D. dissertation, INSA de Toulouse, 2017.

\bibitem{amadeo2014named}
M.~Amadeo, C.~Campolo, A.~Iera, and A.~Molinaro, ``Named data networking for
  iot: An architectural perspective,'' in \emph{Networks and Communications
  (EuCNC), 2014 European Conference on}.\hskip 1em plus 0.5em minus 0.4em\relax
  IEEE, 2014, pp. 1--5.

\bibitem{shi2016edge}
W.~Shi, J.~Cao, Q.~Zhang, Y.~Li, and L.~Xu, ``Edge computing: Vision and
  challenges,'' \emph{IEEE Internet of Things Journal}, vol.~3, no.~5, pp.
  637--646, 2016.

\bibitem{gedeon2019fog}
J.~Gedeon, F.~Brandherm, R.~Egert, T.~Grube, and M.~M{\"u}hlh{\"a}user, ``What
  the fog? edge computing revisited: Promises, applications and future
  challenges,'' \emph{IEEE Access}, vol.~7, pp. 152\,847--152\,878, 2019.

\bibitem{sarddar2018refinement}
D.~Sarddar, S.~Barman, P.~Sen, and R.~Pandit, ``Refinement of resource
  management in fog computing aspect of qos,'' \emph{International Journal of
  Grid and Distributed Computing}, vol.~11, no.~5, pp. 29--44, 2018.

\bibitem{iorga2018fog}
M.~Iorga, L.~Feldman, R.~Barton, M.~J. Martin, N.~S. Goren, and C.~Mahmoudi,
  ``Fog computing conceptual model,'' \emph{NIST: National Institute of
  Standards and Technology}, 2018.

\bibitem{choudhari2018prioritized}
T.~Choudhari, M.~Moh, and T.-S. Moh, ``Prioritized task scheduling in fog
  computing,'' in \emph{Proceedings of the ACMSE 2018 Conference}.\hskip 1em
  plus 0.5em minus 0.4em\relax ACM, 2018, p.~22.

\bibitem{maiti2018qos}
P.~Maiti, J.~Shukla, B.~Sahoo, and A.~K. Turuk, ``Qos-aware fog nodes
  placement,'' in \emph{2018 4th International Conference on Recent Advances in
  Information Technology (RAIT)}.\hskip 1em plus 0.5em minus 0.4em\relax IEEE,
  2018, pp. 1--6.

\bibitem{yousefpour2019all}
A.~Yousefpour, C.~Fung, T.~Nguyen, K.~Kadiyala, F.~Jalali, A.~Niakanlahiji,
  J.~Kong, and J.~P. Jue, ``All one needs to know about fog computing and
  related edge computing paradigms: A complete survey,'' \emph{Journal of
  Systems Architecture}, vol.~98, pp. 289--330, 2019.

\bibitem{chiang2017clarifying}
M.~Chiang, S.~Ha, F.~Risso, T.~Zhang, and I.~Chih-Lin, ``Clarifying fog
  computing and networking: 10 questions and answers,'' \emph{IEEE
  Communications Magazine}, vol.~55, no.~4, pp. 18--20, 2017.

\bibitem{aazam2018deploying}
M.~Aazam, S.~Zeadally, and K.~A. Harras, ``Deploying fog computing in
  industrial internet of things and industry 4.0,'' \emph{IEEE Transactions on
  Industrial Informatics}, vol.~14, no.~10, pp. 4674--4682, 2018.

\bibitem{tordera2016fog}
E.~M. Tordera, X.~Masip-Bruin, J.~Garcia-Alminana, A.~Jukan, G.-J. Ren, J.~Zhu,
  and J.~Farr{\'e}, ``What is a fog node a tutorial on current concepts towards
  a common definition,'' \emph{arXiv preprint arXiv:1611.09193}, 2016.

\bibitem{musumba2013context}
G.~W. Musumba and H.~O. Nyongesa, ``Context awareness in mobile computing: A
  review,'' \emph{International Journal of Machine Learning and Applications},
  vol.~2, no.~1, p.~5, 2013.

\bibitem{abowd1999towards}
G.~D. Abowd, A.~K. Dey, P.~J. Brown, N.~Davies, M.~Smith, and P.~Steggles,
  ``Towards a better understanding of context and context-awareness,'' in
  \emph{International symposium on handheld and ubiquitous computing}.\hskip
  1em plus 0.5em minus 0.4em\relax Springer, 1999, pp. 304--307.

\bibitem{sanchez2006generic}
L.~Sanchez, J.~Lanza, R.~Olsen, M.~Bauer, and M.~Girod-Genet, ``A generic
  context management framework for personal networking environments,'' in
  \emph{Mobile and Ubiquitous Systems: Networking \& Services, 2006 Third
  Annual International Conference on}.\hskip 1em plus 0.5em minus 0.4em\relax
  IEEE, 2006, pp. 1--8.

\bibitem{kaur2018authentication}
A.~Kaur, G.~Rai, A.~Malik \emph{et~al.}, ``Authentication and context awareness
  access control in internet of things: A review,'' in \emph{2018 8th
  International Conference on Cloud Computing, Data Science \& Engineering
  (Confluence)}.\hskip 1em plus 0.5em minus 0.4em\relax IEEE, 2018, pp.
  630--635.

\bibitem{reddy2019context}
R.~V. Reddy, D.~Murali, and J.~Rajeshwar, ``Context-aware middleware
  architecture for iot-based smart healthcare applications,'' in
  \emph{Innovations in Computer Science and Engineering}.\hskip 1em plus 0.5em
  minus 0.4em\relax Springer, 2019, pp. 557--567.

\bibitem{loske2019context}
M.~Loske, L.~Rothe, and D.~G. Gertler, ``Context-aware authentication:
  State-of-the-art evaluation and adaption to the iiot,'' in \emph{2019 IEEE
  5th World Forum on Internet of Things (WF-IoT)}.\hskip 1em plus 0.5em minus
  0.4em\relax IEEE, 2019, pp. 64--69.

\bibitem{kausar2019mutually}
F.~Kausar, W.~Aman, and D.~Al-Abri, ``Mutually authenticated group key
  management protocol for healthcare iot networks,'' in \emph{Proceedings of
  the Future Technologies Conference}.\hskip 1em plus 0.5em minus 0.4em\relax
  Springer, 2019, pp. 1--12.

\bibitem{anton2019putting}
S.~D. Anton, D.~Fraunholz, C.~Lipps, K.~Alam, and H.~D. Schotten, ``Putting
  things in context: Securing industrial authentication with context
  information,'' \emph{arXiv preprint arXiv:1905.12239}, 2019.

\bibitem{trnka2018using}
M.~Trnka, F.~Rysavy, T.~Cerny, and N.~Stickney, ``Using wi-fi enabled internet
  of things devices for context-aware authentication,'' in \emph{International
  Conference on Information Science and Applications}.\hskip 1em plus 0.5em
  minus 0.4em\relax Springer, 2018, pp. 635--642.

\bibitem{smith2018context}
M.~Smith-Creasey, F.~Albalooshi, and M.~Rajarajan, ``Context awareness for
  improved continuous face authentication on mobile devices,'' in \emph{2018
  IEEE 16th Intl Conf on Dependable, Autonomic and Secure Computing, 16th Intl
  Conf on Pervasive Intelligence and Computing, 4th Intl Conf on Big Data
  Intelligence and Computing and Cyber Science and Technology Congress
  (DASC/PiCom/DataCom/CyberSciTech)}.\hskip 1em plus 0.5em minus 0.4em\relax
  IEEE, 2018, pp. 644--652.

\bibitem{habib2015context}
K.~Habib and W.~Leister, ``Context-aware authentication for the internet of
  things,'' in \emph{Elev. Int. Conf. Auton. Auton. Syst. fined}, 2015, pp.
  134--139.

\bibitem{doukas2012bringing}
C.~Doukas and I.~Maglogiannis, ``Bringing iot and cloud computing towards
  pervasive healthcare,'' in \emph{Innovative Mobile and Internet Services in
  Ubiquitous Computing (IMIS), 2012 Sixth International Conference on}.\hskip
  1em plus 0.5em minus 0.4em\relax IEEE, 2012, pp. 922--926.

\bibitem{zhang2016sharing}
H.~Zhang, Z.~Wang, C.~Scherb, C.~Marxer, J.~Burke, L.~Zhang, and C.~Tschudin,
  ``Sharing mhealth data via named data networking,'' in \emph{Proceedings of
  the 3rd ACM Conference on Information-Centric Networking}.\hskip 1em plus
  0.5em minus 0.4em\relax ACM, 2016, pp. 142--147.

\bibitem{garcia2018air4people}
A.~Garcia-de Prado, G.~Ortiz, J.~Boubeta-Puig, and D.~Corral-Plaza,
  ``Air4people: a smart air quality monitoring and context-aware notification
  system,'' \emph{Journal of Universal Computer Science}, vol.~24, no.~7, pp.
  846--863, 2018.

\bibitem{dobrescu2019context}
R.~Dobrescu, D.~Merezeanu, and S.~Mocanu, ``Context-aware control and
  monitoring system with iot and cloud support,'' \emph{Computers and
  Electronics in Agriculture}, vol. 160, pp. 91--99, 2019.

\bibitem{bacstuug2016proactive}
E.~Ba{\c{s}}tu{\u{g}}, M.~Bennis, and M.~Debbah, ``Proactive caching in 5g
  small cell networks,'' in \emph{Towards 5G: Applications, Requirements and
  Candidate Technologies}.\hskip 1em plus 0.5em minus 0.4em\relax Wiley Online
  Library, 2016, pp. 78--98.

\bibitem{zeydan2016big}
E.~Zeydan, E.~Bastug, M.~Bennis, M.~A. Kader, I.~A. Karatepe, A.~S. Er, and
  M.~Debbah, ``Big data caching for networking: Moving from cloud to edge,''
  \emph{IEEE Communications Magazine}, vol.~54, no.~9, pp. 36--42, 2016.

\bibitem{zhao2018collaborative}
X.~Zhao, P.~Yuan, S.~Tang \emph{et~al.}, ``Collaborative edge caching in
  context-aware device-to-device networks,'' \emph{IEEE Transactions on
  Vehicular Technology}, vol.~67, no.~10, pp. 9583--9596, 2018.

\bibitem{lee2018hierarchical}
J.~Lee and J.~Lee, ``Hierarchical mobile edge computing architecture based on
  context awareness,'' \emph{Applied Sciences}, vol.~8, no.~7, p. 1160, 2018.

\bibitem{afzal2017energy}
B.~Afzal, S.~A. Alvi, G.~A. Shah, and W.~Mahmood, ``Energy efficient context
  aware traffic scheduling for iot applications,'' \emph{Ad Hoc Networks},
  vol.~62, pp. 101--115, 2017.

\bibitem{flores2018evidence}
H.~Flores, V.~Kostakos, S.~Tarkoma, P.~Hui, and Y.~Li, ``Evidence-aware mobile
  cloud architectures,'' in \emph{Mobile Big Data}.\hskip 1em plus 0.5em minus
  0.4em\relax Springer, 2018, pp. 65--84.

\bibitem{tang2018multi}
L.~Tang and S.~He, ``Multi-user computation offloading in mobile edge
  computing: A behavioral perspective,'' \emph{IEEE Network}, vol.~32, no.~1,
  pp. 48--53, 2018.

\bibitem{dhumane2018context}
A.~Dhumane, S.~Guja, S.~Deo, and R.~Prasad, ``Context awareness in iot
  routing,'' in \emph{2018 Fourth International Conference on Computing
  Communication Control and Automation (ICCUBEA)}.\hskip 1em plus 0.5em minus
  0.4em\relax IEEE, 2018, pp. 1--5.

\bibitem{guo2013opportunistic}
B.~Guo, D.~Zhang, Z.~Wang, Z.~Yu, and X.~Zhou, ``Opportunistic iot: Exploring
  the harmonious interaction between human and the internet of things,''
  \emph{Journal of Network and Computer Applications}, vol.~36, no.~6, pp.
  1531--1539, 2013.

\bibitem{musolesi2009car}
M.~Musolesi and C.~Mascolo, ``Car: Context-aware adaptive routing for
  delay-tolerant mobile networks,'' \emph{IEEE Transactions on Mobile
  Computing}, vol.~8, no.~2, pp. 246--260, 2009.

\bibitem{dhumane2016routing}
A.~Dhumane, R.~Prasad, and J.~Prasad, ``Routing issues in internet of things: a
  survey,'' in \emph{Proceedings of the international multiconference of
  engineers and computer scientists}, vol.~1, 2016, pp. 16--18.

\bibitem{tran2016collaborative}
T.~X. Tran, A.~Hajisami, P.~Pandey, and D.~Pompili, ``Collaborative mobile edge
  computing in 5g networks: New paradigms, scenarios, and challenges,''
  \emph{arXiv preprint arXiv:1612.03184}, 2016.

\bibitem{pham2019joint}
X.-Q. Pham, T.-D. Nguyen, V.~Nguyen, and E.-N. Huh, ``Joint node selection and
  resource allocation for task offloading in scalable vehicle-assisted
  multi-access edge computing,'' \emph{Symmetry}, vol.~11, no.~1, p.~58, 2019.

\bibitem{tripathi2017adaptive}
V.~Tripathi, ``Adaptive computation offloading in mobile cloud computing.'' in
  \emph{CLOSER}, 2017, pp. 524--529.

\bibitem{zhang2019deep}
K.~Zhang, Y.~Zhu, S.~Leng, Y.~He, S.~Maharjan, and Y.~Zhang, ``Deep learning
  empowered task offloading for mobile edge computing in urban informatics,''
  \emph{IEEE Internet of Things Journal}, vol.~6, no.~5, pp. 7635--7647, 2019.

\bibitem{tu2017cane}
C.~Tu, H.~Liu, Z.~Liu, and M.~Sun, ``Cane: Context-aware network embedding for
  relation modeling,'' in \emph{Proceedings of the 55th Annual Meeting of the
  Association for Computational Linguistics (Volume 1: Long Papers)}, vol.~1,
  2017, pp. 1722--1731.

\bibitem{hu2014multidimensional}
X.~Hu, X.~Li, E.~Ngai, V.~Leung, and P.~Kruchten, ``Multidimensional
  context-aware social network architecture for mobile crowdsensing,''
  \emph{IEEE Communications Magazine}, vol.~52, no.~6, pp. 78--87, 2014.

\bibitem{alirezaie2017ontology}
M.~Alirezaie, J.~Renoux, U.~K{\"o}ckemann, A.~Kristoffersson, L.~Karlsson,
  E.~Blomqvist, N.~Tsiftes, T.~Voigt, and A.~Loutfi, ``An ontology-based
  context-aware system for smart homes: E-care@ home,'' \emph{Sensors},
  vol.~17, no.~7, p. 1586, 2017.

\bibitem{mingozzi2016semantic}
E.~Mingozzi, G.~Tanganelli, C.~Vallati, B.~Mart{\'\i}nez, I.~Mendia, and
  M.~Gonzalez-Rodriguez, ``Semantic-based context modeling for quality of
  service support in iot platforms,'' in \emph{2016 IEEE 17th International
  Symposium on A World of Wireless, Mobile and Multimedia Networks
  (WoWMoM)}.\hskip 1em plus 0.5em minus 0.4em\relax IEEE, 2016, pp. 1--6.

\bibitem{fortino2017modeling}
G.~Fortino, W.~Russo, C.~Savaglio, M.~Viroli, and M.~Zhou, ``Modeling
  opportunistic iot services in open iot ecosystems.'' in \emph{WOA}, 2017, pp.
  90--95.

\bibitem{casadei2019development}
R.~Casadei, G.~Fortino, D.~Pianini, W.~Russo, C.~Savaglio, and M.~Viroli, ``A
  development approach for collective opportunistic edge-of-things services,''
  \emph{Information Sciences}, vol. 498, pp. 154--169, 2019.

\bibitem{fortino2018towards}
G.~Fortino, C.~Savaglio, C.~E. Palau, J.~S. de~Puga, M.~Ganzha, M.~Paprzycki,
  M.~Montesinos, A.~Liotta, and M.~Llop, ``Towards multi-layer interoperability
  of heterogeneous iot platforms: The inter-iot approach,'' in
  \emph{Integration, interconnection, and interoperability of IoT
  systems}.\hskip 1em plus 0.5em minus 0.4em\relax Springer, 2018, pp.
  199--232.

\bibitem{aloi2017enabling}
G.~Aloi, G.~Caliciuri, G.~Fortino, R.~Gravina, P.~Pace, W.~Russo, and
  C.~Savaglio, ``Enabling iot interoperability through opportunistic
  smartphone-based mobile gateways,'' \emph{Journal of Network and Computer
  Applications}, vol.~81, pp. 74--84, 2017.

\bibitem{casadei2019modelling}
R.~Casadei, G.~Fortino, D.~Pianini, W.~Russo, C.~Savaglio, and M.~Viroli,
  ``Modelling and simulation of opportunistic iot services with aggregate
  computing,'' \emph{Future Generation Computer Systems}, vol.~91, pp.
  252--262, 2019.

\bibitem{chang2018learn}
Z.~Chang, L.~Lei, Z.~Zhou, S.~Mao, and T.~Ristaniemi, ``Learn to cache: Machine
  learning for network edge caching in the big data era,'' \emph{IEEE Wireless
  Communications}, vol.~25, no.~3, pp. 28--35, 2018.

\bibitem{pascoal2020mobile}
R.~Pascoal, A.~D. Almeida, and R.~C. Sofia, ``Mobile pervasive augmented
  reality systems—mpars: The role of user preferences in the perceived
  quality of experience in outdoor applications,'' \emph{ACM Transactions on
  Internet Technology (TOIT)}, vol.~20, no.~1, pp. 1--17, 2020.

\bibitem{wattenhofer2004xtc}
R.~Wattenhofer and A.~Zollinger, ``Xtc: A practical topology control algorithm
  for ad-hoc networks,'' in \emph{18th International Parallel and Distributed
  Processing Symposium, 2004. Proceedings.}\hskip 1em plus 0.5em minus
  0.4em\relax IEEE, 2004, p. 216.

\bibitem{maheshwari2018scalability}
S.~Maheshwari, D.~Raychaudhuri, I.~Seskar, and F.~Bronzino, ``Scalability and
  performance evaluation of edge cloud systems for latency constrained
  applications,'' in \emph{2018 IEEE/ACM Symposium on Edge Computing
  (SEC)}.\hskip 1em plus 0.5em minus 0.4em\relax IEEE, 2018, pp. 286--299.

\bibitem{luo2015opportunistic}
J.~Luo, J.~Hu, D.~Wu, and R.~Li, ``Opportunistic routing algorithm for relay
  node selection in wireless sensor networks,'' \emph{IEEE Transactions on
  Industrial Informatics}, vol.~11, no.~1, pp. 112--121, 2015.

\bibitem{han2009reference}
G.~Han, D.~Choi, and W.~Lim, ``Reference node placement and selection algorithm
  based on trilateration for indoor sensor networks,'' \emph{Wireless
  Communications and Mobile Computing}, vol.~9, no.~8, pp. 1017--1027, 2009.

\bibitem{zou2009latency}
F.~Zou, Z.~Zhang, and W.~Wu, ``Latency-bounded minimum influential node
  selection in social networks,'' in \emph{International Conference on Wireless
  Algorithms, Systems, and Applications}.\hskip 1em plus 0.5em minus
  0.4em\relax Springer, 2009, pp. 519--526.

\bibitem{subhlok1999automatic}
J.~Subhlok, P.~Lieu, and B.~Lowekamp, ``Automatic node selection for high
  performance applications on networks,'' \emph{ACM SIGPLAN Notices}, vol.~34,
  no.~8, pp. 163--172, 1999.

\bibitem{zhao2015cooperative}
T.~Zhao, S.~Zhou, X.~Guo, Y.~Zhao, and Z.~Niu, ``A cooperative scheduling
  scheme of local cloud and internet cloud for delay-aware mobile cloud
  computing,'' in \emph{2015 IEEE Globecom Workshops (GC Wkshps)}.\hskip 1em
  plus 0.5em minus 0.4em\relax IEEE, 2015, pp. 1--6.

\bibitem{wang2016edge}
Y.~Wang and L.~Qiu, ``Edge selection-based low complexity detection scheme for
  scma system,'' in \emph{2016 IEEE 84th Vehicular Technology Conference
  (VTC-Fall)}.\hskip 1em plus 0.5em minus 0.4em\relax IEEE, 2016, pp. 1--5.

\end{thebibliography}

\end{document}